# Resolving the ultrafast dynamics of charge carriers in nanocomposites

J. Barreto, T. Roger and A. Kaplan[*]

Nanoscale Physics Research Laboratory, School of Physics and Astronomy

University of Birmingham, Edgbaston, Birmingham, B15 2TT (United Kingdom)

Here we describe an optical method to determine the dynamics of optically excited carriers in nanostructured composite samples. By combining pump-probe time-resolved reflectivity with scattering measurements, we extract the characteristic times for charge carrier evolution. We use the 3D Maxwell-Garnett formulae, modified to include the Drude optical response, to model the results. The method, applied to hydrogenated amorphous silicon containing crystalline silicon nanoparticles, showed that the recombination times in the nanocrystals and in the matrix were ~4.9 ps and ~22 ps, respectively. The charge transfer time between the crystals and the matrix was ~4 ps.

The time domain evolution of free carriers determines the most important features of electro-optical devices: speed, efficiency, and transient non-linear optical behavior[1]. Optical pump-probe spectroscopy has proved itself as the best tool for studying ultra-fast carrier dynamics in homogeneous materials[2]. However, in nanocomposites, pump-probe measurements are unable to differentiate the response of the inclusions from that of their host[3]. Nanocomposites offer the prospect of enabling the development of a multitude of devices and applications[4], but our understanding of these systems is fundamentally limited by the lack of a clear picture of their electro-optical properties. The potential of nanocomposites, materials composed of different phases or elements, one of which has nanoscale dimensions or patterns, was described two decades ago[5]. Yet, owing to the inherent complexity, our comprehension of the characteristic behaviors of such systems is scarcely supported by theoretical models[6] and usually must rely on direct experimental results. Unsurprisingly, there have only been limited attempts to resolve the dynamics in nanostructured composite materials, and as a result, little is known about their optical excitation and decay channels[3,7,8]. In general, determination of the overall charge dynamics requires knowledge of the charge dynamics in both phases of the nanocomposite.

---

[*] Corresponding author: a.kaplan.1@bham.ac.uk

Optical pump-probe techniques are exceptionally good for characterizing the charge dynamics of materials because of their extreme time resolution and the ability to tune the amount of interaction/excitation induced. Nevertheless, measurements are limited by the spatial resolution given by the laser spot size. In nanocomposites, averaging of the optical response over the spot area masks the individual contribution of the nanoparticles.

Here we present an experimental scheme (Figure 1) that enables us to discern the optical response of nanocrystals from their host by a combination of pump-probe time-resolved reflectometry and scattering. The former provides spatially-averaged information about the effective dielectric function, while the latter depends on the difference between dielectric functions of the individual components. Thus, provided that a model of the excitation dependence of the dielectric functions is known, this method allows one to reconstruct independently the evolution of the excitation in both components.

We demonstrate the effectiveness of our technique by studying a nanocrystalline hydrogenated silicon film (nc-Si:H) film containing silicon nanocrystals (Si-nc) and extracting the excitation, decay, and transfer times of optically-pumped carriers in both the embedded nanocrystals and the amorphous matrix.

The system chosen for investigation consists of 500-nm-thick layers of nc-Si:H containing Si-nc. The layers were grown on a crystalline silicon substrate covered by ~200 nm of silicon oxide. Using Raman and X-Ray Diffraction (XRD) measurements, we have estimated that about 35% of the total layer volume is occupied by Si-nc with a mean diameter of ⟨a⟩ = 6 nm. The film growth methods are similar to those described elsewhere[9].

The time resolved optical response of the samples is characterized by a modified pump-probe configuration (Figure 1). With the arrival of a high-intensity "pump" pulse, a density of free carriers, $N(t)$, is generated and their evolution, unknown in the time domain *a priori*, is studied using a second "probe" pulse. By controlling the delay between both pulses, the optical properties of the sample can be analyzed at different instants following excitation. In this study, a Coherent laser system was used, which delivers ~50 fs pulses of up to 3 mJ power at 1 kHz repetition centered at 790 nm. The laser beam was split by a pellicle into the pump and the probe with a corresponding power ratio of 100:1. The probe was delayed by a computer-controlled moving retroreflector. The pump fluence was adjusted using an attenuator based on a combination of near Brewster angle reflectance with a rotatable linear polarizer. The probe and pump were focused to ~50 and ~250 μm spot diameters, respectively. The probe beam angle of incidence was 70° relative to the sample surface normal, while the angle of the pump was 60°. Measurements are taken using a biased silicon photodetector and a Stanford Research SR830 Lock-in amplifier. Further information on the experimental setup may be found elsewhere[10]. Reflectivity was measured at the specular configuration, while the scattering was measured at a 30° scattering angle. No detectable change in the reflectivity and scattering was observed on the substrate over the range of fluences used in this study, $F$=0.4-1.4 mJ/cm$^2$. Nevertheless, measurements performed on samples containing a very low volume fraction (< 0.5 %) of Si-nc yield no noticeable response.

The transient reflectivity, $R(t)$, of a material has been traditionally used to estimate the time evolution of the dielectric function, $\varepsilon$[2,11]:

$$R(t) = \left|\frac{1-\sqrt{\varepsilon(t)}}{1+\sqrt{\varepsilon(t)}}\right|^2 \tag{1}$$

while the dielectric function of a material containing free electrons ($N$) is given by the Drude model:

$$\varepsilon = \varepsilon^0 - \left(\frac{Ne^2}{\varepsilon_0 m^* m_e}\right)^2 \frac{1}{(\omega^2 + i\gamma\omega)} \tag{2}$$

where $\varepsilon^0$ represents the contribution from the ion core background of the sample, and $N$ represents the electron-hole plasma concentration while $e$, $m_e$, and $m^*$ are the electron charge, free electron mass, and optical mass, respectively. The dissipation term $\gamma$ relates to the probability of transition to other available states in the system and $\omega$ is the frequency of the probe. All of these parameters are well known from previous studies of bulk amorphous[12] and crystalline silicon[13,14]. In contrast, for composite materials like the samples used in this work, an effective dielectric function can be described by the Maxwell-Garnett mixing rule[15]:

$$\varepsilon_{eff} = \varepsilon_m + 3f\varepsilon_m \frac{\varepsilon_{nc} - \varepsilon_m}{\varepsilon_{nc} + 2\varepsilon_m - f(\varepsilon_{nc} - \varepsilon_m)} \tag{3}$$

where $\varepsilon_{nc}$ is used to approximate the dielectric function of Si-nc as spherical particles occupying a fractional volume, $f$, surrounded by a homogeneous matrix of hydrogenated amorphous silicon with a dielectric function of $\varepsilon_m$. According to eq. 2, the transient behavior of $\varepsilon_{eff}$ depends, therefore, on both the carrier concentrations in the nanocrystals, $N_{nc}(t)$, and in the matrix, $N_m(t)$. Given the difference between the matrix band gap (~1.7 eV) and the incoming pump energy (1.55 eV), we can assume that the initial excitation takes place predominantly in the Si-nc. Once excited, the free carriers can leak from nanocrystal states into matrix defect states. However, the carriers' spatial redistribution might have little effect on the reflectance, which represents overall average contribution rather than the local carrier environment. Thus, the transient reflectivity cannot be unambiguously interpreted without resorting to additional assumptions or measurements.

On the other hand, through their corresponding dielectric constants, $\varepsilon_{nc}$ and $\varepsilon_m$, both concentration terms, $N_{nc}$ and $N_m$, appear in the description of the scattering efficiency, $Q_{scat}$[16]:

$$Q_{scat} = \frac{8}{3}\left(\frac{2\pi \langle a \rangle}{\lambda}\right)^4 \left|\frac{\varepsilon_{nc}-\varepsilon_m}{\varepsilon_{nc}+\varepsilon_m}\right|^2 \qquad (4)$$

which can be estimated from the measurement of the scattered light intensity, $I_s$:

$$I_S = K I_i Q_{scat} \qquad (5)$$

where $I_i$ corresponds to the probing light intensity and $K$ is a geometrical factor related to the collection efficiency of the light scattered from the irradiated volume.

Hence, by combining the results from reflectivity and scattering, the contributions of the corresponding dielectric functions of the matrix and the nanocrystals can be obtained and related hereafter to the concentration of carriers in each component of the composite.

The results from a traditional pump-probe reflectivity measurement can be seen in Figure 2(a). In comparison, Figure 2(b) shows the time-resolved scattered light intensity. In both cases, the data have been normalized to the reflectivity (and scattering) values before the arrival of the pump, i.e. $\Delta R/R_0 = (R(t)-R_0)/R_0$ (and $\Delta I_S/I_0 = (I_S(t)-I_0)/I_0$). Both signals show an immediate change with the arrival of the pump; remarkably, while $\Delta R$ becomes negative, $\Delta I_S$ increases. Furthermore, it can be seen that the reflectivity decay lasts longer, implying the contribution from two different relaxation mechanisms: a fast one, lasting a few picoseconds (on the timescale of the scattering total decay), and a slow one, in the range of tens of picoseconds. The absence of a slow component in $\Delta I_S$ suggests that the difference between the dielectric functions of the Si-nc and the matrix reaches an undetectable level within the duration of the fast decay process.

Thermal effects are not expected to have a significant contribution because the fractional change of the lattice temperature is insignificant at the pumping level we use in this experiment[13]. It is also usually a relatively slow process developing within a few tens of picosecond, which is longer than the main optical response change investigated in this work.

Using the above equations, the transient fractional reflectivity, $\Delta R/R_0$, and scattering, $\Delta I_S(t)/I_0$ can be reconstructed iteratively (see Figure 2c and d) by finding the best fit to the experimental data in terms of the carrier densities, $N_{nc}(t)$ and $N_m(t)$. The extracted profiles of $N_{nc}(t)$ and $N_m(t)$ can be reasonably well described by the combination of a rising and a decaying exponential function as shown in Figure 2e. The main constants extracted by the fitting procedure are summarized in Table I.

The retrieved transient concentration functions, $N_{nc}(t)$ and $N_m(t)$, provide insights into the carriers dynamics not accessible by other methods. We observe that the excitation of the Si-nc ($\tau_{exc}$ ~85 fs) is almost simultaneous with the pump pulse duration. On the other hand, the maximum concentration of free carriers in the matrix is reached a few picoseconds later, in correlation with the Si-nc decay time ($\tau_{dec}$). Thus we propose (Figure 2f) that the decay process in the Si-nc consists of two channels: recombination within the crystals, measured by the time $\tau_{rec\_nc}$, and charge transfer to the matrix, $\tau_{tran}$, described by a time which approximates the rise time in the matrix. The recombination time can be approximated by $\tau_{rec\_nc} \approx \tau_{dec} \tau_{tran}/(\tau_{tran} - \tau_{dec})$ = 4.9 ps. This estimation is close to the 3 ps surface state decay times measured in nanocrystalline silicon films[7]. Furthermore, we note that the carrier decay time in the matrix, $\tau_{rec\_m}$, of 22 ps is similar to that observed in bulk amorphous silicon,[12] which is attributed to multiparticle recombination.

The self-consistency of our approach was tested by analyzing the probe reflectivity and the scattering intensity as a function of the fluence at the fixed relative delay time of ~100 fs between the pump and the probe. The measured signals and calculated curves are shown in Figure 3. In addition, Figure 3 shows the calculated relative scattering efficiency, $Y_{scat} = Q_{scat}/(Q_{scat} + Q_{abs})$, where $Q_{abs}$ is the absorption efficiency. It is notable that the absorption dominates the scattering significantly. Thus, scattering can only be observed from Si-nc located close to the sample surface since absorption prevails deeper in the sample layer. The comparison of the scattering and absorption probabilities also implies that the contribution of multi-scattered photons to the total scattering is insignificant as statistically, after a single scattering event, scattered photons are absorbed rather than scattered again.

In summary, we have presented a method aimed to resolve the ambiguity inherent in the traditional methods of optical pump-probe spectroscopy applied for composite materials. We demonstrate that the combination of transient reflectivity and scattering measurements allows deconvolution of the evolution dynamics of free carriers in individual components of a composite. The applicability of the method has been demonstrated by measuring the fractional change of reflectivity and scattering on samples of crystalline silicon nanoparticles embedded in an nanocrystalline hydrogenated silicon matrix. Combining a Maxwell-Garnett approximation with the Drude model, we have extracted the transient curves of the carrier concentration in the nanocrystals and the matrix. The method allowed us to establish the characteristic times of the carrier excitation, recombination, and transfer between the matrix and the nanocrystals. We envision that this method will complement and extend traditional pump-probe spectroscopy techniques to carrier dynamics studies in composite and non-uniform materials.

The Coherent laser system used in this research was obtained through the Birmingham Science City project: Creating and Characterizing Next Generation Advanced Materials, supported by Advantage West Midlands (AWM) and funded in part by the European Regional Development Fund (ERDF). J. Barreto acknowledges Advantage West Midlands for an SCRA fellowship. We thank EPSRC for the financial support and Sirica DC for the samples and advice on sample characterization. Finally, we would like to thank I. Yurkevich for the inspiring discussions.

Table I. Characteristic times for excitation and decay of nc-Si:H films containing silicon nanocrystals. The values have been obtained by fitting the datasets in Figure 2a and 2b.

| Component | Si-nc | nc-Si:H (matrix) |
|---|---|---|
| Excitation time, $\tau_{exc}$ [fs] | 85 ± 10 | - |
| Decay time, $\tau_{dec}$ [ps] | 2.2 ± 0.2 | - |
| Transfer time, $\tau_{tran}$ [ps] | 4.0 ± 0.7 | |
| Recombination time, $\tau_{rec}$ [ps] | 4.9 ± 0.5 | 22 ± 4 |
| $N$ at 0.4 mJ/cm$^2$ [$10^{20}$ cm$^{-3}$] | 8.1 ± 0.1 | 1.7 ± 0.1 |
| $N$ at 1.4 mJ/cm$^2$ [$10^{20}$ cm$^{-3}$] | 27.0 ± 0.1 | 2.6 ± 0.1 |

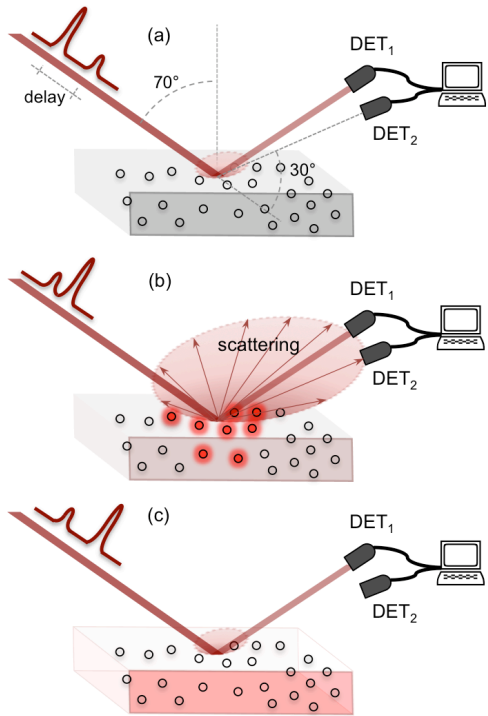

FIG 1. (Color online) Schematic representation of combined pump-probe reflectivity and scattering measurements. $D_1$ and $D_2$ represent detectors that have been oriented to measure the reflected and scattered components, respectively, of the probe beam. Top, middle, and bottom panels correspond to the relative delay times of the probe arriving before, shortly after, and long after the pump beam, respectively. Top panel, (a): the dielectric functions of matrix and nanocrystals are approximately equal, $\varepsilon_m \approx \varepsilon_{nc}$; there are not yet any excited carriers, $N_{nc} = N_m = 0$; the reflected and scattered components are used as a reference. Middle panel, (b): $\varepsilon_m > \varepsilon_{nc}$; $N_{nc} > N_m$; the pumping of the nanocrystals induces a negative change in the reflectivity and increases the scattering. Bottom panel, (c): $\varepsilon_m \approx \varepsilon_{nc}$, $N_{nc} \approx N_m \neq 0$; the excited carriers leak into the matrix and the scattering diminishes as the carrier concentration in the sample becomes uniform. The balance between transfer and recombination processes in both matrix and nanocrystals decreases the average concentration, and the reflectivity recovers toward its initial level.

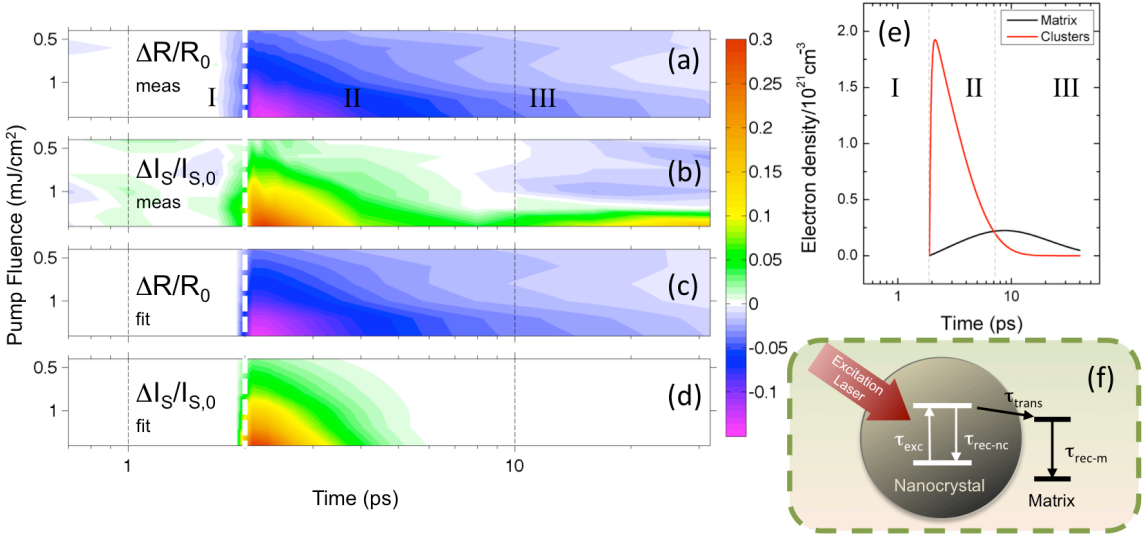

FIG 2. Time-resolved optical response and dynamics of the free carriers in the nanocomposite. (a)-(b) Measured and (c)-(d) Calculated 2D color maps of the reflectivity $\Delta R/R_0$, and the scattering $\Delta I/I_0$ change. The x-axis represents relative delay time (on logarithmic scale) between the pump and the probe and the y-axis corresponds to the fluence of the pump laser. The arrival of the pump is highlighted with a white dashed line at about 2 ps. The regions denoted I, II, and III correspond to the conditions described in Figure 1(a), (b), and (c), respectively. (e) Carrier dynamics in the nanocrystals and the matrix reconstructed from (c)-(d) at a fluence of 1 mJ cm$^{-2}$. (f) Schematic representation of the photo-excited carrier dynamics in the material; carriers are pumped in the nanocrystals with a response time $\tau_{exc}$, excited carriers can recombine within the Si-nc through $\tau_{rec\text{-}nc}$ or leak into matrix states through $\tau_{trans}$; carriers in matrix states recombine within $\tau_{rec\text{-}m}$ (see Table I).

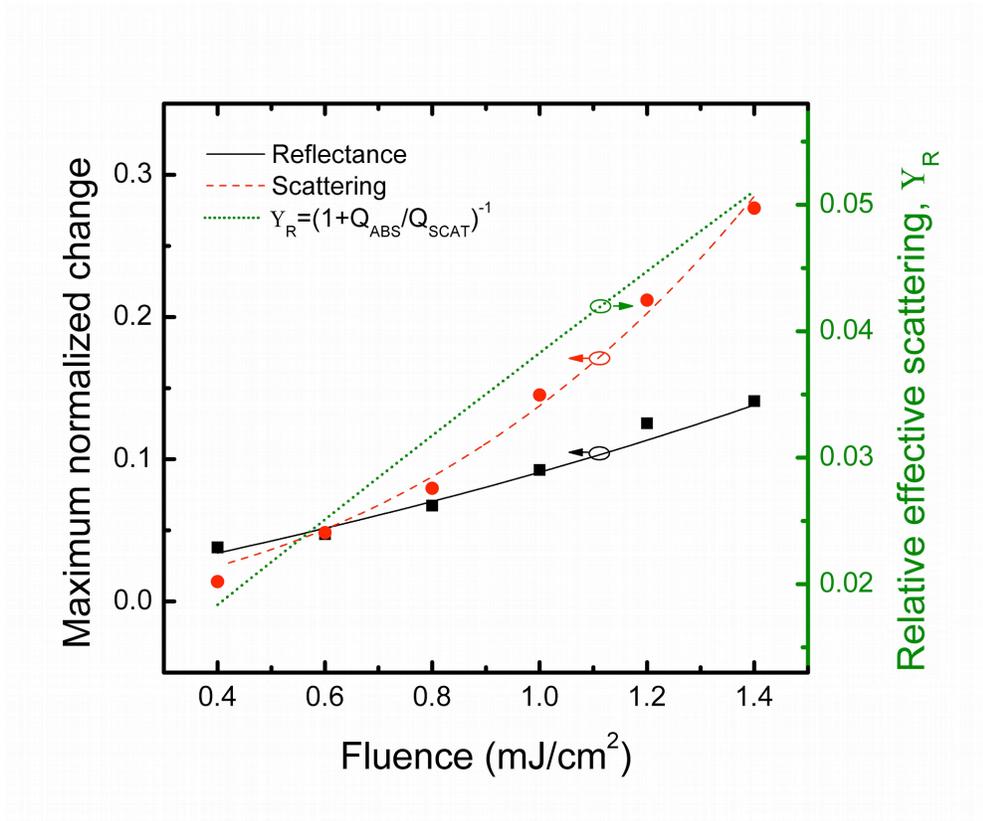

FIG 3. (Color online) Dependence of the optical properties of nc-Si:H films on laser fluence. The left y-axis shows maximum reflectance and scattering change, the right axis corresponds to the relative scattering efficiency as a function of the laser fluence. The dots show the experimental data and the lines represent the calculations.